\documentclass[a4paper,11pt]{article}
\pdfoutput=1
\usepackage{jheppub}
\usepackage[T1]{fontenc}
\usepackage[utf8]{inputenc}
\usepackage{braket}
\usepackage{subcaption}

\graphicspath{ {./plots/} }

\renewcommand{\Re}{\operatorname{Re}}

\newcommand{\mul}{\mu^\text{loop}}
\newcommand{\E}{\operatorname{\mathbb{E}}}
\newcommand{\tr}{\operatorname{tr}}
\newcommand{\nl}{N}
\newcommand{\nw}{N_\text{w}}
\newcommand{\G}[1]{\Gamma\left(#1\right)}
\newcommand{\ac}{ \alpha_{\mathbb S} }

\title{\boldmath New Recipes for Brownian Loop Soups}

\author{Valentino F.\ Foit}
\author{Matthew Kleban}
\affiliation{Center for Cosmology and Particle Physics, New York University, 726 Broadway, New York, NY 10003, USA}

\emailAdd{foit@nyu.edu}
\emailAdd{kleban@nyu.edu}

\abstract{
We define a large new class of conformal primary operators in the  ensemble of Brownian loops in two dimensions known as the ``Brownian loop soup,'' and compute their correlation functions analytically and in closed form.  The loop soup is a conformally invariant statistical ensemble with central charge $c = 2 \lambda$, where $\lambda > 0$ is the intensity of the soup. Previous work identified exponentials of the layering operator $e^{i \beta N(z)}$ as primary operators. Each Brownian loop was assigned $\pm 1$ randomly, and $N(z)$ was defined to be the sum of these numbers over all loops that encircle the point $z$. These exponential operators then have conformal dimension ${\frac{\lambda}{10}}(1 - \cos \beta)$. Here we generalize this procedure by assigning a more general random value to each loop. The  operator $e^{i \beta N(z)}$ remains primary with conformal dimension $\frac {\lambda}{10}(1 - \phi(\beta))$, where $\phi(\beta)$ is the characteristic function of the probability distribution used to assign random values to each loop. Using recent results we compute in closed form the exact two-point functions in the upper half-plane and four-point functions in the full plane of this very general class of operators. These correlation functions depend analytically on the parameters $\lambda, \beta_i, z_i$, and on the characteristic function $\phi(\beta)$. They satisfy the conformal Ward identities and are crossing symmetric. As in previous work, the conformal block expansion of the four-point function reveals the existence of additional and as-yet uncharacterized conformal primary operators.
}

\begin{document} 
\maketitle
\flushbottom

\section{Introduction}

Conformally invariant theories in two dimensions are of great interest, both due to their wide applicability to various physical systems and to the high degree of analytic control conformal symmetry provides \cite{Belavin:1984vu, Cardy:2008jc, Ginsparg:1988ui, DiFrancesco:639405}. Some conformal theories can be defined via a local Lagrangian density, others are defined by a random ensemble or statistical system. One as yet little explored theory in the latter class is the so-called Brownian Loop Soup (BLS) \cite{Lawler2004}. Very recently, the discovery of a new technique allowed for the four-point function of certain operators in the BLS on the plane (and the two-point function in the upper half-plane) to be calculated analytically and in closed form \cite{Camia:2019ots}. While the precise relation of the BLS to other, better-known conformal field theories (CFTs) remains obscure, this result should provide a key tool in exploring these connections.

As the name suggests, the ingredients of the soup are Brownian loops -- random Brownian motions that return to the same point after some specified Brownian ``time'' $t$, so that they form (generally self-intersecting) closed loops.
Due to the well-known fact that the standard deviation of the excursion in Brownian motion scales as $\sqrt{t}$, the mean area of such a loop is proportional to $\sqrt{t}^2 = t$.

The BLS ensemble is a ``Poissonian gas'' of these random loops, with locations $z$ chosen uniformly randomly (i.e.\ with measure $d^2 z$) in some two dimensional domain, and with time durations chosen randomly with measure $d t/t^2$. Since $t$ has dimensions of area, the scale invariant measure on a single loop is
\begin{align} \label{BLSmeasure}
    d \mul = \frac{1}{2 \pi} \, d^2 z \, \frac{d t}{t^2} \, d \mu^\text{br}(z,t),
\end{align}
where $d \mu^\text{br}(z,t)$ is the measure for a Brownian loop at $z$ with duration $t$ (known as the complex Brownian bridge measure).

The partition function of the full BLS with intensity $\lambda>0$ is 
\begin{align} \label{partfunc}
    Z = 1+ \sum_{n=1}^{\infty} \frac{\lambda^n}{n!} \prod_{k=1}^n \int d\mul = \exp\left( \lambda \mul \right).
\end{align}
Each term in the sum over $n$ corresponds to a configuration of exactly $n$ loops, weighted by $\lambda^n$, and divided by $n!$ (to account for identical configurations). The product and integral is over all possible shapes of the $n$ loops.

The BLS turns out to be more than just scale invariant -- it is fully conformally invariant in a very strict sense \cite{Lawler2004} and has central charge\footnote{Note that in some references the central charge was incorrectly given as $c=\lambda$.} \cite{Camia:2015ewa}
\begin{align} \label{c}
    c = 2 \lambda.
\end{align}
This formula for the central charge can be derived in an illuminating way by considering a massless, free scalar field (a Gaussian free field) in two spatial dimensions, which is well-known to be a CFT with $c = 1$. The logarithm of the partition function for the free field satisfies \cite{Camia:2015ewa}
\begin{align} \label{GFF}
    \mul = 2 \log Z_\text{free boson}.
\end{align}
Because the central charge is additive in the logarithm of $Z$ (for instance, the central charge of two non-interacting theories is the sum of their central charges) the relation \eqref{GFF} establishes the relation \eqref{c}. Furthermore, it shows that the BLS with intensity $\lambda$ can be thought of as ``$2 \lambda$ copies of a free field.'' This suggests that for half-integer $\lambda$, the theory may have special or simplifying features \cite{camia2015nonbacktracking, lejan2010, vandebrug2018}.

Another implication of \eqref{c} is that the theory cannot be a unitary (or reflection positive) CFT for generic $\lambda < 1/2$. The only unitary CFTs with $c < 1$ are the minimal models, and these exist only for a discrete set of possible values of $c$.  Of course, there is no reason for a statistical ensemble such as the BLS to be described by a unitary theory (for instance, critical percolation is non-unitary). However, one sees that the BLS does have some of the ``healthy'' features of unitary CFTs: the central charge is always positive, and the conformal dimensions of all the primary operators we will define and find via the conformal block expansion are positive as well.

\subsection{Background and previous work}

The BLS has no known Lagrangian description (at least for $\lambda \not\in \mathbb Z/2$). However, to characterize it more fully we can attempt to identify physically natural quantities in the theory, and the primary operators that compute them. Indeed, the genesis of this project was \cite{Freivogel:2009rf}, which was attempting to find a description of future infinity of eternal inflation. There, the idea was that phase transitions would produce spherical bubbles that would expand, collide, and overlap. Due to the translation invariance and exponential growth with constant Hubble parameter of de Sitter spacetime, these bubbles would have a scale-invariant distribution of radii and uniformly random center locations -- that is, they would form a scale-invariant distribution rather similar to that of the BLS, but composed of disks (or in higher dimensions, spheres) rather than random loops.

If two or more types of phase transitions are possible, the bubbles at future infinity will have a label attached to them characterizing which type of transition they represent. A simple example is a field theory with a periodic potential, where transitions from any given phase can either shift the field to the minimum to the right or to the left of the initial point. In this model each disk on future infinity can be characterized as $+1$ or $-1$, and the signed sum over all disks overlapping a given point gives the total shift of the field at that point from some fiducial initial value. 

In the ``disk model'' of \cite{Freivogel:2009rf} this corresponds to randomly assigning $\pm 1$ to each disk with equal probability, and then defining a field $N(z) = N_+(z) - N_-(z)$ that counts the overlaps. It turns out that $N(z)$ is a field with dimension zero, and the ``vertex operators'' $e^{i \beta N(z)}$ are conformal primaries. Because $N$ assumes integer values, the dimension of vertex operators in the disk model
\begin{align}
    \Delta(\beta) = \pi \lambda (1 - \cos \beta)
\end{align}
is a periodic function of $\beta$.

As a putative CFT, the disk model of \cite{Freivogel:2009rf} has an apparent flaw. While the authors were able to prove that the disk distribution is translation invariant and invariant under \emph{global} conformal transformations -- rotations, dilitations, and special conformal transformations -- they showed by direct computation that the four-point function of primaries exhibits a pathology, namely a non-analyticity when the fourth operator crosses the circle that connects the other three. This likely indicates that the theory is not locally conformally invariant.\footnote{If so, this is a rare example: a theory invariant under scaling, rotations, and special conformal transformations, but not local conformal transformations.}

In \cite{Camia:2015ewa}, the authors attempted to rectify this deficiency by considering the BLS rather than the disk ensemble of \cite{Freivogel:2009rf}. Because the BLS is provably fully conformally invariant, \cite{Camia:2015ewa} speculated that the correlation functions of the analogous ``vertex'' operators would be meromorphic and that the theory would define a proper CFT. In fact, \cite{Camia:2015ewa} identified two distinct classes of operators: ``layering'' operators analogous to those defined in \cite{Freivogel:2009rf}, where each loop is assigned $\pm 1$ with equal probability and the operator $\nl(z)$ counts the signed sum over all loops with outer boundaries that encircle (or layer) the point $z$ (Fig.\ \ref{loops}), and ``winding operators'' where the loop is assigned an orientation and $\nw(z)$ counts the sum of the winding numbers of all loops at $z$ (where loops that do not encircle $z$ have winding number zero). Either of these can be given a physical interpretation similar to the one described above: if the Brownian loops themselves (for the winding operator) or their outer boundaries (for the layering operator) represent a domain wall or charged object across which some quantity (like the value of a scalar field, or the electric field) changes discontinuously, then these operators are counting the value of that field at the point $z$ (this is also known as a height model).
The layering and winding vertex operators in the infinite intensity limit have been studied in \cite{camia2019brownian, camia2020limit}.

\subsection{Relation to known CFTs}

Statistical models that can be described in terms of random loops include the so-called random-cluster models \cite{Grimmett_2006}, the Ising model, the $q$-state Potts model, and the $O(n)$ vector model.
In fact, the $O(n)$ model in the limit $n \to 0$ can be used to determine  the conformally invariant ensemble of single self-avoiding loops \cite{Gamsa_2006}. Through a uniqueness theorem of Werner \cite{werner2005conformally}, this same ensemble also describes the outer boundaries of conformally invariant Brownian loops, which our layering number vertex operators are sensitive to.  This is the tool that made the results of \cite{Camia:2019ots} for the four-point function in the plane possible to obtain.

The Conformal Loop Ensembles $\text{CLE}_\kappa$ with parameter $\frac{8}{3} < \kappa \le 4$ describe the scaling limit of loop cluster boundaries in various critical statistical models \cite{sheffield2009, 10.2307/23350642}.
For intensities $0 < \lambda \le \frac{1}{2}$, the BLS is related to the $\text{CLE}_\kappa$ through $\lambda = \frac{(3 \kappa -8)(6 - \kappa)}{4 \kappa}$.

\section{Summary and results}

In this note we will present a  natural generalization of the BLS operators defined in \cite{Camia:2015ewa}. Rather than assigning $\pm 1$ to each loop, we will assign some more general random labels. For instance, we will consider assigning a real number drawn from a probability distribution function (PDF) such as a normal distribution, with support on a continuous interval rather than on the integers.  In this case, the operator $N(z)$ in a particular ensemble of loops will take a value equal to the sum over all these random variables for the loops that encircle the point $z$ (and zero for any loop that does not). Another generalization is to assign a vector, and then study correlations of the operator $e^{i \boldsymbol{\beta} \cdot \mathbf{N}(z)}$, where $\boldsymbol{\beta}$ is a vector of parameters. This provides a large new class of conformal primaries and a beautifully simple and universal formula for their conformal dimensions.

\subsection{Correlation functions}

\begin{figure}[t]
    \centering
    \begin{subfigure}[t]{0.4\textwidth}
        \includegraphics[width=\textwidth]{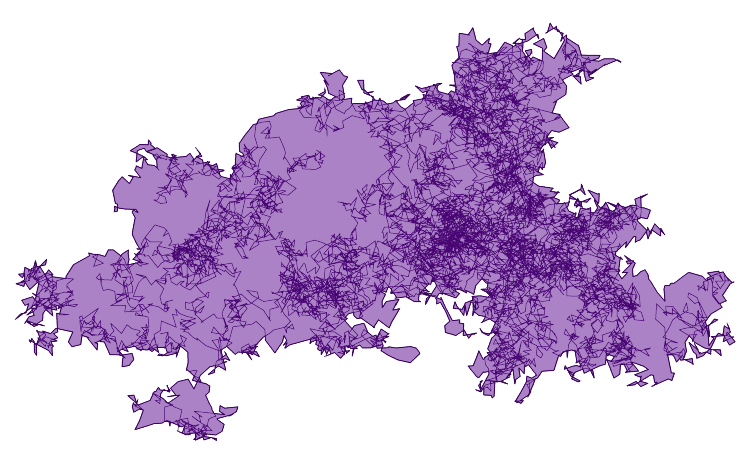}
        \caption{A Brownian loop (dark violet) with its interior shaded (pale violet).}
        \label{1loop}
    \end{subfigure}~
    \begin{subfigure}[t]{0.5\textwidth}
        \includegraphics[width=\textwidth]{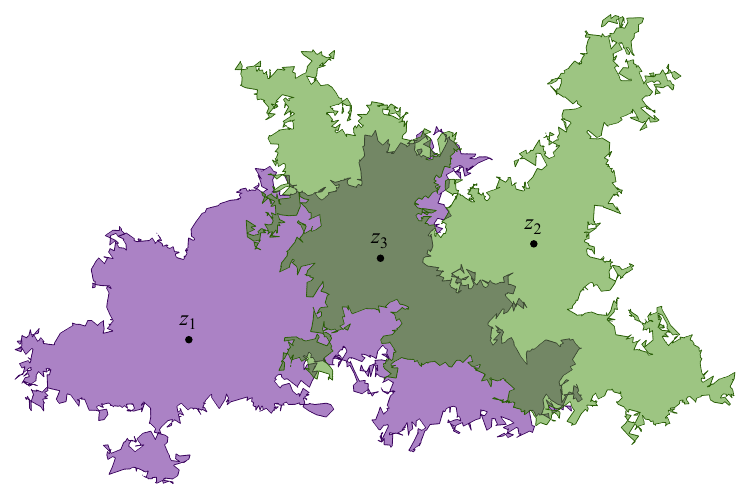}
        \caption{The interiors of two  Brownian loops (pale violet and pale green) and their area of intersection (dark green).}
        \label{2loops}
    \end{subfigure}
    \caption{The Brownian loop soup is a weighted sum over the number of loops and all possible configurations of each. (\subref{1loop}) A single loop and its interior. In the class of models considered here, in every configuration the $i$th loop is assigned a random value $x_i$ drawn from some specified distribution (in the model previously considered this distribution was taken to be $\pm1$ with equal probability). Correlation functions are computed by summing over loop configurations and these random distributions. (\subref{2loops}) If the violet loop is assigned random value $x_1$ and green $x_2$, the operator $e^{i \beta_1 N(z_1)}e^{i \beta_2 N(z_2)}e^{i \beta_3 N(z_3)}$ equals $e^{i \beta_1 x_1}e^{i \beta_2 x_2} e^{i \beta_3 (x_1+x_2)}$ for this configuration.}
    \label{loops}
\end{figure}

We consider $n$-point correlation functions
\begin{align}
	\Braket{\prod_{j=1}^n \mathcal{O}_{\beta_j}(z_j)}
\end{align}
of $\mathcal{O}_\beta(z) \sim e^{i \beta N(z)}$ which are exponentials of $i$ times a real number $\beta$ times numbering operator $N(z)$.
These correlation functions are a generalized version of those computed in \cite{Camia:2019ots}.
Each loop in a given collection of Brownian loops is assigned a random value chosen from some distribution.  Then for each loop that encircles the point $z$, $N(z)$ receives an additive contribution. The additive contribution is either the random value assigned to the loop times 1 (for the layering operator that we will focus on in the rest of the paper, denoted $N(z)$), or the random value times the winding number $\nw(z)$ of the loop around $z$, cf.\ Appendix \ref{winding}. It is the choice of distribution for the random variable that generalizes the work of \cite{Camia:2015ewa,Camia:2019ots} (where in this language the random variable was $\pm1$ with equal probability), cf.\ Fig.\ \ref{2loops}.

The $\mathcal{O}_\beta(z)$ are conformal primary operators with novel conformal scaling dimensions.
They transform under a conformal map $f: D \rightarrow D'$ with $z_j' = f(z_j)$ as
\begin{align}
	\Braket{\prod_{j} \mathcal{O}_{\beta_j}(z_j')}_{D'} = \prod_{j}|f'(z_j)|^{-2 \Delta(\beta_j)} \Braket{\prod_{j} \mathcal{O}_{\beta_j}(z_j)}_D.
\end{align}
For exponentials of the layering operator we find (Sec.\ \ref{laydimsec})
\begin{align} \label{laydim}
	\Delta(\beta) =  \frac{\lambda}{10}(1 - \phi(\beta)),
\end{align}
and for exponentials of the winding operator (Appendix \ref{winding})
\begin{align}
	\Delta_\text{w}(\beta) =  \frac{\lambda}{2 \pi^2}\sum_{m=1}^{\infty}\frac{1}{m^2} (1 - \phi(m \beta) ).
\end{align}
The function $\phi$ is the characteristic function of the random variable that multiples the layering or winding number of each loop
\begin{align}
	\phi(\beta) = \E\left[e^{i \beta X}\right],
\end{align}
where $\E[\,\cdot\,]$ denotes expectation value with respect to an even probability distribution. Characteristic functions exist for arbitrary random objects, such as random vectors, random matrices, and random functions and the form of \eqref{laydim} remains valid.

In the upper half-plane, we compute the one- and two-point functions of exponentials of layering operators (Sec.\ \ref{uhp})
\begin{align} 
    \Braket{\tilde {\mathcal O}_{ \beta_1}(z_1) }_{\mathbb{H}}
    =&  |z_1-\overline z_1|^{-2\Delta_1}
\end{align}
and
\begin{align}
\begin{split}
&\Braket{\tilde {\mathcal O}_{ \beta_1}(z_1) \tilde {\mathcal O}_{\beta_2} (z_2)}_{\mathbb{H}} \\
    = &|z_1-z_2|^{-2(\Delta_1+\Delta_2-\Delta_{12})} |z_1-\overline z_2|^{2(\Delta_1+\Delta_2-\Delta_{12})} |z_1-\overline z_1|^{-2\Delta_1}  |z_2-\overline z_2|^{-2\Delta_2} \\
    &\times \exp\left[ -(\Delta_1+\Delta_2- \Delta_{12})(1-\sigma) {}_3 F_{2}\left(1,1,\frac{4}{3};2,\frac{5}{3};  1-\sigma \right) \right],
\end{split}
\end{align}
where $\Delta_i \equiv \Delta(\beta_i), ~ \Delta_{ij} \equiv \Delta(\beta_i+\beta_j)$, and $\sigma = \frac{|z_1-z_2|^2}{|z_1-\overline z_2|^2}$. The tilde in $\tilde {\mathcal O}_{ \beta}(z) \sim e^{i \beta N(z)}$ denotes a normalization appropriate for the upper half plane, chosen so that the coefficient of the one-point function is unity.

In the full plane, correlations functions of layering vertex operators vanish unless the ``charge conservation condition''
\begin{align} \label{cc}
	\sum_{j=1}^{n} \beta_j= \frac{2\pi}{b}k
\end{align}
is imposed, where $k$ is an integer if the random variable $X$ is given by a lattice distribution (the period $b$ is the lattice spacing), and $k=0$ otherwise.  When this condition holds, the two-, three-, and four-point functions of the normalized exponentials of layering operators are
\begin{align} 
   \Braket{ {\mathcal O}_{ \beta_1}(z_1)  {\mathcal O}_{\beta_2} (z_2)}_\mathbb{C} &=  |z_1 - z_2|^{-4\Delta_1}, \label{full2pt}\\
     \Braket{\prod_{i=1}^3 {\mathcal O}_{\beta_i}(z_i)}_\mathbb{C} &= 
    |z_1-z_2|^{-2(\Delta_1+\Delta_2-\Delta_{3})}
    |z_1- z_3|^{-2(\Delta_1+\Delta_3-\Delta_{2})}
    |z_2- z_3|^{-2(\Delta_2+\Delta_3-\Delta_{1})}, \label{full3pt}
\end{align}
and
\begin{align}
\begin{split}
    \Braket{\prod_{i=1}^4 {\mathcal O}_{\beta_i}(z_i)}_\mathbb{C} 
    &= \exp\left[- 2 A(x) \left( \sum_{i=1}^4 \Delta_i - \sum_{j=2}^4 \Delta_{1 j} \right) \right]
    \left|\frac{z_{13} z_{24}}{z_{12}z_{34}}\right|^{-  2\Delta_{12}}
   \left|\frac{z_{13} z_{24}}{z_{14}z_{23}}\right|^{-  2\Delta_{14}} \\
   & \quad \times \left |\frac{z_{12} z_{14}}{z_{24}}\right|^{-2 \Delta_1}
   \left |\frac{z_{12} z_{23}}{z_{13}}\right|^{- 2\Delta_2} 
   \left|\frac{z_{23} z_{34}}{z_{24}}\right|^{- 2\Delta_3} 
   \left|\frac{z_{14} z_{34}}{z_{13}}\right|^{-2 \Delta_4},
\end{split}
\end{align}
where $z_{ij} \equiv z_i - z_j$, $x \equiv \frac{z_{12} z_{34}}{z_{13} z_{24}}$ is the cross-ratio, and
\begin{align}
\begin{split}
    A(x) &= \frac{1}{4} \left[
        x ~ {}_3 F_{2}\left(1,1,\frac{4}{3};2,\frac{5}{3}; x\right)
        +\overline x ~ {}_3 F_{2}\left(1,1,\frac{4}{3};2,\frac{5}{3}; 
        \overline x\right)  \right] \\
    & - 6 \mu | x (1-x) |^{\frac{2}{3}} \left|
    {}_2 F_{1}\left(\frac{2}{3},1;\frac{4}{3}; 
    x \right)\right|^2
\end{split}
\end{align}
with
\begin{align}
    \mu = \frac{2^{\frac{1}{3}} \pi^2}{3 \sqrt{3} \G{\frac 1 6}^2 \G{\frac 4 3}^2}. 
\end{align}
The normalization of ${\mathcal O}_{ \beta}(z) \sim e^{i \beta N(z)}$ is again chosen so that the coefficient of the two-point function is unity.
Remarkably, with this choice the three-point functions coefficients are unity as well \cite{Camia:2019ots}.

\section{New scaling dimensions of layering operators} \label{laydimsec}

As we explained in the introduction, the disc model of \cite{Freivogel:2009rf} and the loop model of \cite{Camia:2015ewa} were defined by randomly assigning a binary variable $\pm 1$ to each disc or loop in the ensemble. 
We now generalize this procedure by assigning arbitrary random objects to the loops in the ensemble and compute new, previously unknown conformal dimensions. From this point on we only consider loops in the BLS and exponentials of the layering number which are conformal primaries.

One of the main results of \cite{Camia:2015ewa} was an explicit formula for $n$-point functions of exponential layering operators in the BLS in terms of their weights and charges $\beta_j$
\begin{align} \label{npointlay}
    \Braket{\prod_{j=1}^n
    e^{i \beta_j N(z_j)}
    }_D = &\prod_{\substack{S \subseteq \{z_1, \ldots , z_n \} \\ S \neq \emptyset}} \exp \left[
    - \lambda \,  \alpha_D(S|S^c) \left(1 - \cos\sum_{i \in I_S} \beta_i  \right) \right].
\end{align}
The product on the right-hand side is over all nonempty subsets $S \subseteq \{ z_1, \ldots, z_n \}$  and $I_S$ denotes the set of indices $i$ corresponding to the points $z_i \in S$. The $\alpha_D(S|S^c)$ are the weights of the sets of loops that encircle the points in $S$ but not those in $S^c$ according to the Brownian loop measure
\begin{align} \label{weightdef}
\begin{split}
    \alpha_D(S|S^c) &= \mul(\gamma: \operatorname{diam}(\gamma) >\delta,
    \gamma \subseteq D, z \in \overline \gamma ~ \forall z \in S, z \not\in \overline \gamma ~ \forall z \in S^c ).
\end{split}
\end{align}
The loops are contained in the domain $D$, which in this paper is the upper half-plane $\mathbb{H}$ or the full plane $\mathbb{C}$. It is understood that no subscript refers to the full plane, i.e.\ $\alpha \equiv \alpha_\mathbb{C}$ and $\braket{\,\cdot\,} \equiv \braket{\,\cdot\,}_\mathbb{C}$.

In the next section we derive the general scaling dimensions for layering operators $\nl$ and in Sec.\ \ref{npointlay_gen} we generalize \eqref{npointlay}.
Specific examples of correlation functions involving random scalar, vector, matrix and random function regularizations are given in Sec.\ \ref{examples}.
In Sec.\ \ref{applications} we apply those results to correlation functions in the upper half plane, as well as the full plane.
It is possible to naturally extend this generalization to the winding operator which is given in Appendix \ref{winding}. It is noteworthy, however, that the results in the upper half plane, as well as the four-point function in the full plane (Sec.\ \ref{applications}) are not known for winding operators due to our ignorance regarding the appropriate weights of loops winding around certain sets of points.

\subsection{Scaling dimensions} \label{scalingdims}

We now derive the scaling behavior of the one-point function of exponential layering operators. Due to conformal invariance, the one-point function vanishes in the plane. However, evaluating it in a disk of radius $R$ (or alternatively, with a cutoff on the diameter of the loops, defined as the largest distance between any pair of points on the loop) reveals the scaling dimensions of the conformal primaries. This computation is meant to be illustrative and to showcase the origin of new scaling dimensions. A more general approach is shown in Sec.\ \ref{npointlay_gen}.

The idea considered in this paper is to assign a random object $X$ to every loop in the ensemble. The operator $\nl(z)$ is equal to the sum over all these objects for each loop with an outer boundary that encircles the point $z$.
The one-point function is then
\begin{align}
\begin{split}
    \Braket{ e^{i \beta \nl(z)}}_{\delta,R} &= \E\left[e^{i \beta \nl(z)}\right]\\
&= \sum_{k=0}^\infty \E\left[ e^{i \beta \nl(z)} | \mathcal{L}_k \right] P(\mathcal{L}_k),
\end{split}
\end{align}
where
\begin{align}
    \mathcal{L}_k = \{ \eta : | \left\{ \gamma \in \eta : z \in \bar{\gamma}, \delta \le \operatorname{diam}(\gamma) < R \right\} | = k \}
\end{align}
is the ensemble of $k$ loops that cover the point $z$. The cutoffs $\delta$ and $R$ on the smallest and largest diameters of loops which are necessary to render the result finite,
and $P(\mathcal{L}_k) = \frac{(\lambda \alpha)^k }{k!} e^{- \lambda \alpha}$ is Poissonian since every loop in the ensemble is independent.

The expectation value $\E[\,\cdot\,]$ is taken with respect to a probability distribution. If the mean of the distribution is non-zero we can simply replace $X$ with $X - \E[X]$, which corresponds to absorbing the effect of the non-zero mean into the normalization of the exponential operators.

Therefore, we will assume the mean vanishes and we also require that the distribution is even. As we will see, the latter requirement guarantees that the conformal dimensions are real. It can also be motivated by considering the BLS on a sphere, where there is no distinction between the inside and the outside of a loop and the charge conservation condition (\ref{cc}) is required for consistency. One of the results of \cite{Camia:2019ots} was that the weights that contribute to all correlation functions (in the plane and on the sphere) are symmetric under $\beta \to -\beta$, so that only the even part of the distribution  contributes. Therefore without loss of generality we can require the distribution to be even.

First we evaluate the expectation value
\begin{align}
    \E\left[ e^{i \beta \nl(z)} | \mathcal{L}_k \right] &= \E\left[ e^{i \beta \sum^* X_\gamma} | \mathcal{L}_k \right] \\
    &= \E\left[ e^{i \beta X} \right]^k.
\end{align}
The second equality follows from the fact that all loops are independent of each other; we used the notation
\begin{align}
    \nl(z) = \sum_{\substack{\gamma \in \eta, z \in \bar\gamma \\ \delta \le \operatorname{diam}(\gamma)<R}} X_\gamma = {\sum}^* X_\gamma.
\end{align}
The function
\begin{align} \label{phi}
    \phi(\beta) \equiv \E\left[e^{i \beta X}\right]
\end{align}
is  the characteristic function of the probability distribution for $X$, the properties of which we analyze in detail in Sec.\ \ref{characteristic}.
This allows us to evaluate the one-point function
\begin{align}
    \Braket{ e^{i \beta \nl(z)}}_{\delta,R} = \sum_{k=0}^\infty \phi(\beta)^k \frac{(\lambda \alpha)^k}{k!} e^{-\lambda \alpha}= e^{-\lambda \alpha (1-\phi(\beta))},
\end{align}
where $\alpha \equiv \alpha_{\mathbb{C}}(z) = \mul(\gamma: z \in \bar{\gamma}, \delta \le \operatorname{diam}(\gamma) < R)$ is the weight that a single loop covers any point, with a short-distance cutoff $\delta$ and a long-distance cutoff $R$.
It was shown in \cite{Camia:2015ewa} that $\alpha = \frac{1}{5} \log \frac{R}{\delta}$ and so we obtain
\begin{align} \label{1pt}
    \Braket{ e^{i \beta \nl(z)}}_{\delta,R} =  \left( \frac{R}{\delta} \right)^{- \frac{\lambda}{5} (1 - \phi(\beta))} = \left( \frac{R}{\delta} \right)^{-2 \Delta(\beta)}.
\end{align}
This vanishes in the limit $\delta \to 0$, but the $\delta$ dependence can be absorbed into a multiplicative normalization of $e^{i \beta \nl}$ to define an operator $\mathcal O_\beta$ that remains finite when $\delta$ is taken to zero (this is the analogue of wave-function renormalization in quantum field theory). The one-point function \eqref{1pt} also vanishes as $R \to \infty$ because it does not satisfy the charge conservation condition \eqref{cc}. In the rest of the paper, we will consider only correlations functions that satisfy \eqref{cc} and can therefore define them in the limit $\delta \to 0, R \to \infty$.

We can read off the scaling dimension easily from the behavior under rescaling $R$. The conformal dimension is evidently
\begin{align} \label{confdim}
    \Delta(\beta) = \frac{\lambda}{10} (1 - \phi(\beta)),
\end{align}
correctly reproducing a result of \cite{Camia:2015ewa} when $\phi(\beta) = \cos(\beta)$.

\subsection{Correlation functions} \label{npointlay_gen}

To confirm and to generalize \eqref{confdim} to the $n$-point functions, one can follow the derivation of the general correlation function in \cite{Camia:2015ewa}. The generalization of \eqref{npointlay} is simply
\begin{align} \label{layeringcorr}
\begin{split}
    \Braket{\prod_{j=1}^n
    e^{i \beta_j \nl(z_j)}}
    &= \E\left[ {e^{i \sum_{j=1}^n \beta_j \nl(z_j)}} \right] \\
    &= \prod_{\substack{S \subseteq \{z_1, \ldots , z_n \} \\ S \neq \emptyset}} \exp \left[
    - \lambda \, \alpha(S|S^c) \left(1 - \phi\left(\sum_{i \in I_S} \beta_i \right) \right) \right].
\end{split}
\end{align}
As we discuss in the next section, the values of $\beta_j$ must obey the so called ``charge conservation condition''
\begin{align}
    \sum_{j=1}^n \beta_j = 0.
\end{align}
Additional conditions are permissable for periodic characteristic functions with
\begin{align}
    \sum_{j=1}^n \beta_j = \frac{2 \pi}{b} k, ~ k \in \mathbb{Z}, ~ b \in \mathbb{R}
\end{align}
which are discussed in Sec.\ \ref{lattice}.

\subsection{Characteristic functions} \label{characteristic}

The expectation values we encountered \eqref{phi} are the Fourier transforms of probability distributions and are known as characteristic functions. We now discuss their properties.

For a real-valued random variable the characteristic function
\begin{align}
	\phi(\beta) = \int d x ~ p(x) e^{i \beta x}
\end{align}
always exists and is uniformly continuous. It is
\begin{subequations} \label{phiproperties}
\begin{align}
    \text{bounded} \quad &|\phi(\beta)| \le 1, \label{bounded}\\
    \text{normalized} \quad &\phi(0) = 1, \label{normalized} \\
    \text{Hermitian} \quad &\phi(-\beta) = \overline{\phi(\beta)}. \label{herm}
\end{align}
\end{subequations}
The characteristic function is real if and only if the probability distribution is even.
We also see that an even probability density function produces an even characteristic function.
As we mentioned earlier, only even distributions define sensible CFTs in the plane and on the sphere.

We note that properties \eqref{bounded} and \eqref{normalized} ensure that the conformal dimensions are non-negative and vanish in at least one point:
\begin{align}
\begin{split}
    \Delta(\beta) &\ge 0, \\
    \Delta(0) &= 0.
\end{split}
\end{align}

We can now easily generalize from scalar variables and discuss random vectors, random matrices, and random functions. The properties \eqref{phiproperties} apply to characteristic functions of these distributions as well.

Multivariate random variables can be interpreted as random vectors, which may have random lengths, as well as random orientations. For a $d$-dimensional random complex vector $\boldsymbol{\beta, X} \in \mathbb{C}^d$ we have
\begin{align}
	\phi(\boldsymbol{\beta}) = \E\left[ \exp \left(i \Re\left( \boldsymbol{\beta}^\dagger \boldsymbol{X} \right) \right) \right],
\end{align}
where $\dagger$ denotes the conjugate transpose.
Similarly, one can consider random matrices $B, X \in \mathbb{R}^{d \times d'}$ and their characteristic function
\begin{align}
	\phi(B) = \E\left[\exp\left(i \tr(B^\dagger X)\right)\right],
\end{align}
where $\tr(\cdot)$ denotes the trace.
Random functions can be thought of as infinite-dimensional random vectors and we are dealing with a characteristic functional. For square integrable functions $x,\beta$ on some domain $V$, one may compute
\begin{align}
	\phi[\beta] = \E\left[\exp\left(
	i \int_V dt ~ \overline{\beta(t)} x(t)
	\right)\right].
\end{align}
There are not many random functions for which this expression can be evaluated analytically. Some cases, such as random Gaussian functions which is the continuum limit of the multivariate normal distribution, are known.

\subsection{Examples} \label{examples}

\begin{figure}[t]
    \centering
    \begin{subfigure}[b]{0.45\textwidth}
        \includegraphics[width=\textwidth]{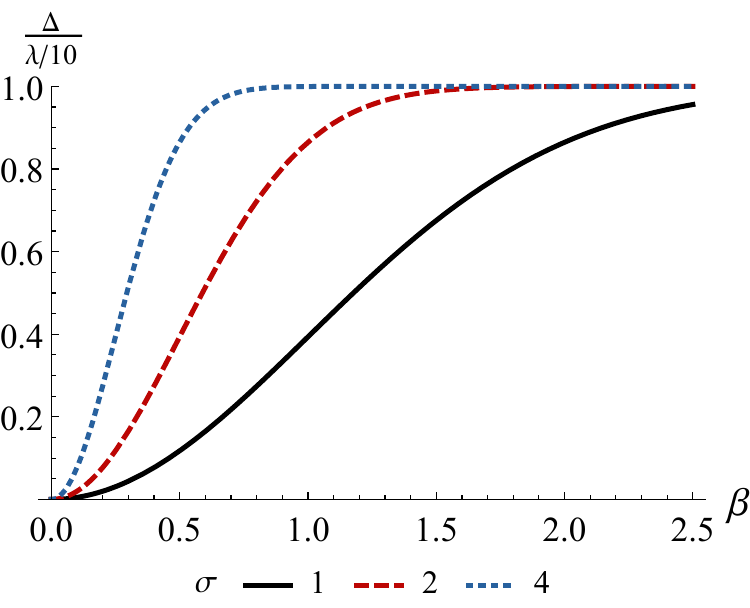}
        \caption{Conformal dimensions of the vertex operators when the distribution is a single Gaussian random variable, for several values of the standard deviation.}
        \label{phi1}
    \end{subfigure}~ ~
    \begin{subfigure}[b]{0.45\textwidth}
        \includegraphics[width=\textwidth]{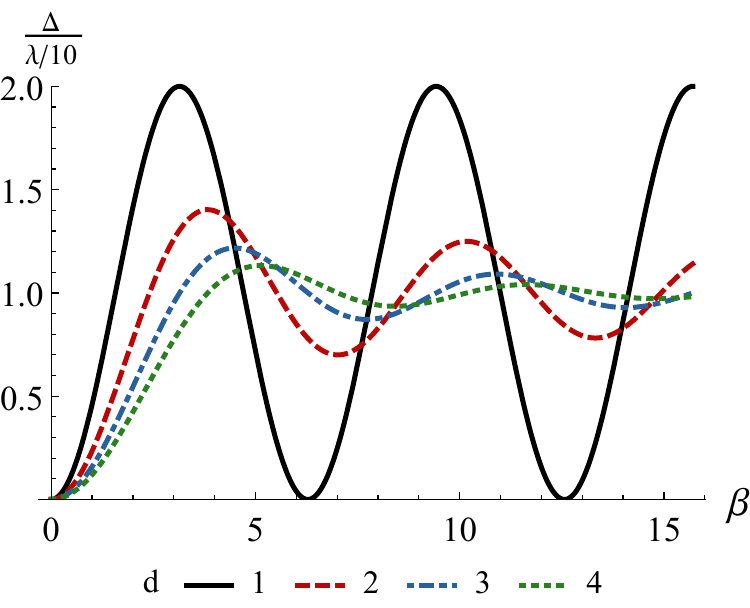}
        \caption{Conformal dimensions of the vertex operators when the distribution is random vectors of unit length in several different dimensions $d$.}
        \label{phi2}
    \end{subfigure}
    \caption{Two examples of conformal dimensions $\Delta(\beta)$ are shown as a function of $\beta$.}
    \label{phis}
\end{figure}

A simple example is a Gaussian random variable with zero mean
\begin{align}
    p(x) = \frac{1}{\sqrt{2\pi}\sigma} e^{-\frac{x^2}{2\sigma^2}},
\end{align}
that has the characteristic function
\begin{align}
\begin{split}
    \phi(\beta) &= \E\left[ e^{i \beta X} \right] \\
    &= e^{-\frac{1}{2} \sigma^2 \beta^2}.
\end{split}
\end{align}
The corresponding conformal dimensions
\begin{align}
    \Delta(\beta) = \frac{\lambda}{10}\left(1 - e^{- \frac{1}{2} \sigma^2 \beta^2}\right)
\end{align}
are plotted in Fig.\ \ref{phi1}.

Next, let us compute the characteristic function of a real $d$-dimensional random vector $\boldsymbol{X} \in \mathbb{R}^d$ of unit length $|\boldsymbol{X}|=1$.
For fixed $\boldsymbol{\beta} \in \mathbb{R}^d$ we average the random vector over all orientations of $\boldsymbol{X}$.
Due to the $O(d)$ symmetry of the problem, the characteristic function is only a function of $\beta = |\boldsymbol{\beta}|$.
One obtains
\begin{align} \label{randvec}
\begin{split}
    \phi(\beta) &= \E\left[ e^{i \boldsymbol{\beta} \cdot \boldsymbol{X}} \right] \\
    &= \frac{1}{S_{d-1}} \int d \Omega ~ e^{i \boldsymbol{\beta} \cdot \boldsymbol{X}} \\
    &= {}_0F_1\left(;\frac{d}{2};-\frac{\beta^2}{4}\right),
\end{split}
\end{align}
where we denote the surface of the $d-1$-sphere
\begin{align}
    S_{d-1} = \int d \Omega = \frac{2 \pi^{\frac{d}{2}}}{\Gamma\left(\frac{d}{2}\right)}.
\end{align}
${}_0F_1$ is known as the confluent hypergeometric function\footnote{It is e.g.\ related to the Bessel function of the first kind via
\begin{align*}
	{}_0F_1(;a;-b) = \Gamma(a)b^{\frac{1-a}{2}} J_{a-1}(2\sqrt{b}).
\end{align*}}
and it has the properties
\begin{align}
\begin{split}
    {}_0F_1(;a;0) &= 1, \\
    \lim\limits_{x \rightarrow \infty}{}_0F_1(;a;-x) &= 0 \quad \text{if } a>1.
\end{split}
\end{align}
The case $a = \frac{d}{2} = 1$ is of particular interest and discussed in Sec.\ \ref{bern}.
The conformal dimensions
\begin{align}
    \Delta(\beta) = \frac{\lambda}{10}\left(1 - {}_0F_1\left(; \frac{d}{2}; -\frac{\beta^2}{4} \right)  \right)
\end{align}
are plotted in Fig.\ \ref{phi2}.

\subsubsection{Periodic characteristic functions} \label{lattice}

We now study a class of scalar probability distributions of special interest.
A lattice distribution is a distribution for which the probabilities are non-zero only for points that form a subset of a lattice, $P(X \in a + b \mathbb{Z}) = 1$ for real $a,b$.
A characteristic function is periodic with real period if and only if it belongs to a lattice distribution which has the origin as a lattice point, i.e.\ $a = 0$ \cite{CM_1956-1958__13__76_0}.
In other words, periodic characteristic functions belong to scalar probability density functions of the form
\begin{align}
	p(x) = \sum_{n=-\infty}^\infty p_n \delta(x - b n) \quad \text{with} \quad \sum_{n=-\infty}^\infty p_n = 1.
\end{align}
The associated characteristic functions
\begin{align}
    \phi(\beta) = \sum_{n=-\infty}^\infty p_n e^{i \beta b n}
\end{align}
always exist, are single valued and analytic around the origin, and are periodic with real period $2\pi/b$. When $\beta$ obeys the ``charge conservation'' condition
\begin{align}
    \beta^* = \frac{2 \pi}{b} k, ~ k \in \mathbb{Z}
\end{align}
we have $\phi(\beta^*) =  \phi(0) = 1$.
Real, even, periodic characteristic functions belong to even lattice distributions with $p_n = p_{-n}$.
Note that characteristic functions with imaginary periods exist, but, for our purposes, are not of interest.

\subsubsection{The Bernoulli distribution} \label{bern}

It is now apparent that the conformal dimensions of the exponential of the layering number from \cite{Camia:2015ewa,Camia:2019ots} are a special case of the more general prescription.
As we discussed in the introduction, the authors assigned  $\pm 1$ to each loop with equal probability and defined $\nl(z) = N_+(z) - N_-(z)$.
For random variables distributed according to the Bernoulli distribution
\begin{align}
    p(x) = \frac{1}{2}\left[\delta(x-1)+\delta(x+1)\right],
\end{align}
the characteristic function is
\begin{align}
    \phi(\beta) = \cos\beta.
\end{align}
Alternatively, this result can be understood as a one-dimensional random vector of unit length. From \eqref{randvec} and $d=1$
\begin{align}
    {}_0F_1\left(;\frac{1}{2};-\frac{\beta^2}{4}\right) = \cos\beta.
\end{align}
We thus reproduce the conformal scaling dimension of the exponential of the layering operator from the previous work
\begin{align}
	\Delta(\beta) = \frac{\lambda}{10} (1- \cos\beta),
\end{align}
which has period $2\pi$, as shown in Fig.\ \ref{phi2}.

\section{Applications} \label{applications}

We  now compute correlation functions of our generalized operators in the BLS using the results of \cite{Camia:2019ots}. We show the results for the one- and two-point function in the upper half plane, as well as for the four-point function in the full plane. With the latter we obtain the generalized three-point function coefficients in the BLS.

We denote the conformal dimensions by
\begin{align}
\begin{split}
    \Delta_j &=\frac{\lambda}{10} (1-\phi(\beta_j)) \\
    \text{and}\quad \Delta_{i j} &= \frac{\lambda}{10}(1-\phi(\beta_i+\beta_j)),
    \end{split}
\end{align}
where $\phi$ is a characteristic function drawn from the set of functions we described in the previous section.

\subsection{In the upper half-plane} \label{uhp}

We normalize the exponentials of layering operators in the upper half-plane $\mathbb{H}$ by
\begin{align}\label{halfnorm}
    \tilde{\mathcal O}_\beta(z) \equiv \left({2 \delta e^{-5 \hat{\alpha}}  }\right)^{-2\Delta(\beta)}e^{i \beta \nl(z)},
\end{align}
where $\hat{\alpha} = \mul(\gamma:  \text{diam}(\gamma) > 1, \gamma \subseteq \mathbb{H}, z_1 \in \bar{\gamma})$ is the constant weight of the loops with diameter greater than or equal to $1$ contained in $\mathbb{H}$ and winding around the point $z=i$.
The $n$-point correlation functions are defined by the limit
\begin{align} \label{uhplimit}
    \Braket{\prod_{j=1}^n \tilde{\mathcal{O}}_{\beta_j}(z_j) } = \lim_{\delta \rightarrow 0} \frac{\Braket{\prod_{j=1}^n e^{i \beta_j \nl(z_j)}}}
    {\left( 2 \delta e^{-5 \hat{\alpha}}
    \right)^{2 \sum_{j=1}^n \Delta_j}}.
\end{align}

The one-point function is
\begin{align} \label{onept}
    \Braket{\tilde {\mathcal O}_{ \beta_1}(z_1) }_{\mathbb{H}}
    =&  |z_1-\overline z_1|^{-2\Delta_1}
\end{align}
and the two-point function is
\begin{align}
\begin{split}
&\Braket{\tilde {\mathcal O}_{ \beta_1}(z_1) \tilde {\mathcal O}_{\beta_2} (z_2)}_{\mathbb{H}} \\
    = &|z_1-z_2|^{-2(\Delta_1+\Delta_2-\Delta_{12})} |z_1-\overline z_2|^{2(\Delta_1+\Delta_2-\Delta_{12})} |z_1-\overline z_1|^{-2\Delta_1}  |z_2-\overline z_2|^{-2\Delta_2} \\
    &\times \exp\left[ -(\Delta_1+\Delta_2- \Delta_{12})(1-\sigma) {}_3 F_{2}\left(1,1,\frac{4}{3};2,\frac{5}{3};  1-\sigma \right) \right],
\end{split}
\end{align}
where $\sigma = \frac{ |z_1-z_2|^2 }{  |z_1-\overline z_2|^2 }$. The multiplicative factor in \eqref{halfnorm} was chosen so that \eqref{onept}
is canonically normalized.

\subsection{In the full plane} \label{fp}

The one-point function in the full plane vanishes, and the two- and three-point functions are determined by their scaling dimensions and three-point function coefficients alone, as given in \eqref{full2pt} and \eqref{full3pt}.

\subsubsection{The four-point function}

The first non-trivial correlation function of primary operators in a CFT is the four-point function, which is of the form
\begin{align}
    \Braket{\varphi_1(z_1) \varphi_2(z_2) \varphi_3(z_3) \varphi_4(z_4)} = f(x) \prod_{i=1 <j}^4 z_{i j}^{\Delta/3 - \Delta_i - \Delta_j} \bar{z}_{i j}^{\bar{\Delta}/3 - \bar{\Delta}_i - \bar{\Delta}_j},
\end{align}
where $\varphi_i$ are four not necessarily identical primary operators, $z_{i j} = z_i - z_j, ~ \Delta = \sum_{i=1}^4 \Delta_i$, and $f(x)$ is a function only of the cross ratio
\begin{align}
    x = \frac{z_{12} z_{34}}{z_{13} z_{24}}.
\end{align}

Consider four points $z_1,z_2,z_3,z_4$ and assume in what follows that the letters $i,j,k,\ell \in \{ 1,2,3,4 \}$ are always different. Using \eqref{npointlay}, the four-point function can be written as
\begin{align}
\begin{split} \label{4pt}
\Braket{ \prod_{j=1}^4 e^{i \beta_j N(z_j)} }
= \exp \Bigg[ -\lambda \Bigg(
    &\sum_{i=1}^4 (1-\phi(\beta_i)) \alpha(z_i|z_j, z_k, z_{\ell}) \\
    +& \sum_{\substack{i,j=1\\i<j}}^4(1-\phi(\beta_i+\beta_j)) \alpha(z_i, z_j| z_k, z_{\ell}) \\
    +& \sum_{i=1}^4 (1-\phi(\beta_j+\beta_k+\beta_\ell)) \alpha(z_j,z_k,z_{\ell}|z_i)\\
    +& (1-\phi(\beta_1+\beta_2+\beta_3+\beta_4)) \alpha(z_1,z_2,z_3,z_4)
    \Bigg) \Bigg],
\end{split}
\end{align}
where the weights $\alpha$ of loops encircling points in the full plane are defined in \eqref{weightdef}. Imposing the charge conservation condition and utilizing that the characteristic functions need to be even, the four-point function becomes
\begin{align}
\begin{split} \label{4ptac}
    \Braket{ \prod_{j=1}^4 e^{i \beta_j N(z_j)} }
    = \exp \Bigg[ - \lambda \Bigg(
    &\sum_{i=1}^4(1-\phi(\beta_i)) \ac(z_i|z_j, z_k, z_{\ell}) \\
    + &\sum_{j=2}^4(1-\phi(\beta_1+\beta_j)) \ac(z_1, z_j| z_k, z_{\ell})
    \Bigg) \Bigg],
\end{split}
\end{align}
where we denoted the weights
\begin{align}
    \ac(S | S^c) \equiv \alpha(S | S^c) + \alpha( S^c | S)
\end{align}
for non-empty subsets of points $S \subseteq \{z_1,z_2,z_3,z_4\}$, with $S^c$ denoting the complement of $S$.  For instance, $\ac(z_1 |z_2,z_3,z_4) = \alpha(z_1 |z_2,z_3,z_4) + \alpha(z_2,z_3,z_4 | z_1)$.

A detailed derivation of the weights $\ac$ is given in \cite{Camia:2019ots}. Using the definitions
\begin{align} \label{Ax}
\begin{split}
    A(x) &= \frac{1}{4} \left[
        x ~ {}_3 F_{2}\left(1,1,\frac{4}{3};2,\frac{5}{3}; x\right)
        +\overline x ~ {}_3 F_{2}\left(1,1,\frac{4}{3};2,\frac{5}{3}; 
        \overline x\right)  \right] \\
    & - 6 \mu | x (1-x) |^{\frac{2}{3}} \left|
    {}_2 F_{1}\left(\frac{2}{3},1;\frac{4}{3}; 
    x \right)\right|^2
 \end{split}
 \end{align}
 with
 \begin{align}
     \mu = \frac{2^{\frac{1}{3}} \pi^2}{3 \sqrt{3} \G{\frac 1 6}^2 \G{\frac 4 3}^2} 
 \end{align}
 we obtain the four-point function of exponentials of layering operators in the BLS
\begin{align}
\begin{split} \label{4ptfull}
    \Braket{\prod_{j=1}^4 {\mathcal O}_{\beta_j}(z_j)}
    &= \exp\left[- 2 A(x) \left( \sum_{i=1}^4 \Delta_i - \sum_{j=2}^4 \Delta_{1 j} \right) \right]
    \left|\frac{z_{13} z_{24}}{z_{12}z_{34}}\right|^{-  2\Delta_{12}}
   \left|\frac{z_{13} z_{24}}{z_{14}z_{23}}\right|^{-  2\Delta_{14}} \\
   & \quad \times \left |\frac{z_{12} z_{14}}{z_{24}}\right|^{-2 \Delta_1}
   \left |\frac{z_{12} z_{23}}{z_{13}}\right|^{- 2\Delta_2} 
   \left|\frac{z_{23} z_{34}}{z_{24}}\right|^{- 2\Delta_3} 
   \left|\frac{z_{14} z_{34}}{z_{13}}\right|^{-2 \Delta_4},
\end{split}
\end{align}
where the $\delta \rightarrow 0$ limit has been performed analogously to \eqref{uhplimit} with the normalization
\begin{align}
    \mathcal O_\beta(z) \equiv \left({2 \delta e^{-\frac{\pi}{\sqrt{3}}-5 \hat{\alpha}}  }\right)^{-2\Delta}e^{i \beta N(z)}
    = e^{-\frac{\pi}{\sqrt{3}}} \tilde{\mathcal O}_\beta(z)
\end{align}
chosen such that the two-point function \eqref{full2pt} is canonically normalized.

It is easy to confirm that the four-point function \eqref{4ptfull} satisfies the conformal Ward identities.  Crossing symmetry  implies that \eqref{4ptfull} is invariant under the exchange of any pair of indices, which is guaranteed by the following identities:
\begin{align}
\begin{split}
    A(x) - A(1-x) &= 0 \\
    A(x) - A(1/x) + \log |x| &= 0.
\end{split}
\end{align}

\subsubsection{Conformal block expansion and three-point coefficients}

\begin{figure}[t]
    \centering
    \includegraphics[width=.33\textwidth]{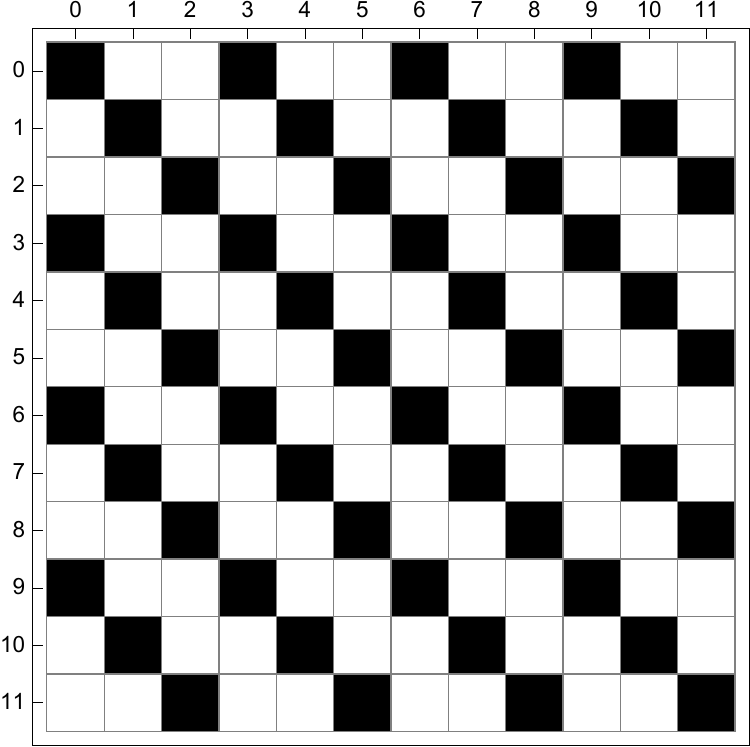}
    \caption{The non-zero products $C_{34}^{(p, p')} C_{12}^{(p,p')}$ are shown for generic scaling dimensions up to order $\mathcal{O}(x^4)$.} 
    \label{matrix}
\end{figure}

The four-point function of a conformal field theory contains information about the three-point function coefficients, as well as the spectrum of primary operators. To obtain this data, one makes use of the operator algebra by performing a conformal block expansion.
We carried out this expansion using the results from the Bernoulli-distribution regularization in previous work. Again, we extend the work to the general case.

By setting $z_1 = \infty, ~z_2 = 1, ~z_3 = x$ and $z_4 = 0$ we define
\begin{align} \label{defG1}
    G^{2 1}_{3 4}(x)  &= \lim_{z_1 \rightarrow \infty} z_1^{2 \Delta_1} \bar{z}_1^{2 \bar{\Delta}_1}
    \Braket{ {\mathcal O}_{\beta_1}(z_1) {\mathcal O}_{\beta_2}(1) {\mathcal O}_{\beta_3}(x) {\mathcal O}_{\beta_4}(0)}_\mathbb{C},
\end{align}
where $\bar{\Delta}_1={\Delta}_1$ in our case (note that later on we will consider operators with spin, $\Delta^{(p,p')} \neq \bar{\Delta}^{(p,p')})$.
We can now proceed to expand the four-point function in Virasoro conformal blocks
\begin{align} \label{confblexp}
    G^{21}_{34}(x) = \sum_{\mathcal{P}} C_{34}^\mathcal{P} C_{12}^\mathcal{P} {\mathcal F}^{21}_{34}(\mathcal{P}|x)  \bar{\mathcal F}^{21}_{34}(\mathcal{P}|\bar x).
\end{align}
The sum over $\mathcal{P}$ runs over all primary operators in the theory, and the $C_{ij}^\mathcal{P}$ are the three-point function coefficients of the operators labeled by $i, j$ with $\mathcal{P}$. Each $\mathcal{P}$ with a non-zero $C$ contributes a term consisting of a holomorphic function times an anti-holomorphic function of the cross-ratio. The Virasoro conformal blocks depend only on $x$, the central charge $c$, and the conformal dimensions $\Delta_i, \Delta_\mathcal{P}$ of the five operators.

By equating \eqref{defG1} and \eqref{confblexp} order by order in $x, \bar{x}$, we can solve for the three-point coefficients $C_{34}^\mathcal{P} C_{12}^\mathcal{P}$. A detailed analogous discussion and derivation is given in \cite{Camia:2019ots}.
We find the following results.
The conformal dimensions of the conformal primary operators under general regularizations are given by
\begin{align}
\begin{split}
    \Delta^{(p,p')} &= \frac{\lambda}{10}(1-\phi(\beta_1+\beta_2)) + \frac{p}{3} \\
    \bar{\Delta}^{(p,p')} &= \frac{\lambda}{10}(1-\phi(\beta_1+\beta_2)) + \frac{p'}{3}.
\end{split}
\end{align}
In Fig.\ \ref{matrix}, we show some non-zero three-point coefficients that appear in expansion \eqref{confblexp}. Note that the series expansion does not terminate at finite order.

The first few terms that appear on the diagonal $C_{34}^{(n,n)} C_{12}^{(n,n)}$ are
\begin{align}
    C_{34}^{(0,0)} C_{12}^{(0,0)} &= 1 \\
    \begin{split}
    C_{34}^{(1,1)} C_{12}^{(1,1)} &= \frac{6}{5} \lambda \mu [1 - \phi(\beta_1) - \phi(\beta_2) - \phi(\beta_3) - \phi(\beta_4) \\
    &+ \phi(\beta_1+\beta_2) + \phi(\beta_1+\beta_3) + \phi(\beta_1+\beta_4) ]
    \end{split}
\end{align}
and for $n \le 3$ we have
\begin{align}
    C_{34}^{(n,n)} C_{12}^{(n,n)} = \frac{1}{n!} \left( C_{34}^{(1,1)} C_{12}^{(1,1)} \right)^n.
\end{align}
The next term has contributions from the hypergeometric function ${}_3 F_2$ from \eqref{Ax} and is significantly more complicated.

The first off-diagonal term is
\begin{align} \label{c03}
     C_{34}^{(0,3)} C_{12}^{(0,3)} = \frac{\lambda}{20} \left[
     \frac{(\phi(\beta_1) - \phi(\beta_2)) (\phi(\beta_3) - \phi(\beta_4))}{1 - \phi(\beta_1+\beta_2)} - \phi(\beta_1 + \beta_3) + \phi(\beta_1 + \beta_4)
     \right].
\end{align}
This term vanishes identically if we identify $\phi(\cdot) \rightarrow \cos(\cdot)$ but does not vanish generally.
The next two terms are
\begin{align}
\begin{split}
    &C_{34}^{(1,4)} C_{12}^{(1,4)} = \frac{\lambda}{20} C_{34}^{(1,1)} C_{12}^{(1,1)} \\
    &\times \left[ \frac{ 3 \lambda (\phi(\beta_1) - \phi(\beta_2)) (\phi(\beta_3) - \phi(\beta_4))}{10 + 3 \lambda (1 -  \phi(\beta_1 + \beta_2))} 
    - \phi(\beta_1 + \beta_3) + \phi(\beta_1 + \beta_4) \right]
\end{split}\\
\begin{split}
    &C_{34}^{(2,5)} C_{12}^{(2,5)} = \frac{\lambda}{40} \left( C_{34}^{(1,1)} C_{12}^{(1,1)} \right)^2 \\
    &\times \left[ \frac{ 3 \lambda (\phi(\beta_1) - \phi(\beta_2)) (\phi(\beta_3) - \phi(\beta_4))}{20 + 3 \lambda (1 -  \phi(\beta_1 + \beta_2))} 
    - \phi(\beta_1 + \beta_3) + \phi(\beta_1 + \beta_4) \right].
\end{split}
\end{align}

While it is straightforward to obtain all other three-point function coefficients, they grow in length rapidly, and we chose not to quote them here. The results for some special cases for which some of those simplify significantly are given again in \cite{Camia:2019ots}.

\section{Other results}

\subsection{Relation to the free boson}

The partition function of the BLS \eqref{partfunc} has been identified with the partition function of a free, massless, real bosonic feld in two Euclidean dimensions 
when the intensity and central charge satisfy $c = 2 \lambda = 1$ \cite{Camia:2015ewa}.
The BLS partition function for general intensity can then be related to the bosonic one through
\begin{align} \label{ff}
	Z(\lambda) = Z(1/2)^{2\lambda}.
\end{align}
This relation is a characteristic of the distribution of the loops themselves, not the random variables we have attached to them, and thus is still valid  here.  Note that one could interpret \eqref{ff} by saying that the BLS with intensity $\lambda$ corresponds to $2 \lambda$ copies of the free field.  However, the operators we consider do not to our knowledge have a simple representation in terms of the free field.

Furthermore, \eqref{layeringcorr} and \eqref{npointwin} show that a similar relation holds for the $n$-point correlation functions. The BLS weights $\alpha$  are  independent of the intensity, and so
\begin{align}
    \Braket{
    \prod_{j=1}^n e^{i \beta_j N(z_j)}}_{D,\lambda} = \Braket{\prod_{j=1}^n e^{i \beta_j N(z_j)}}_{D,\lambda = 1/2}^{2 \lambda}.
\end{align}

\subsection{Free field limit}

The characteristic functions are moment generating
\begin{align}
    \E\left[x^k\right] = \left. \frac{1}{i^k} \frac{d^k}{d \beta^k} \phi(\beta) \right|_{\beta=0}.
\end{align}
It follows that if the $k$th derivate of $\phi$ at 0 does not exist then moments greater or equal than $k$ do not exist.

Earlier we have seen that all sensible characteristic functions for our purpose are even and have the property that $\phi(0) = 1$. This means that if the variance (the second moment) of the distribution exists, the first derivative of $\phi$  vanishes at the origin.  

Given that the variance of the probability distribution exists, there exists a limit in which the correlators in the full plane become those of free field vertex operators.
Consider now taking $\beta_i \rightarrow 0$ and $\lambda \rightarrow \infty$ with the product $\lambda \beta_i^2$ fixed (this limit is discussed in detail in \cite{camia2019brownian}).
The Taylor expansion of the characteristic function is
\begin{align}
    \phi(\beta) = 1 - \frac{1}{2} \E\left[x^2\right] \beta^2 + \mathcal{O}(\beta^4),
\end{align}
so we can express the conformal dimensions in this limit as
\begin{align}
    \lim_{\beta \rightarrow 0, \lambda \rightarrow \infty} \Delta(\beta) = \frac{\lambda}{20} \E\left[x^2\right] \beta^2.
\end{align}
If we define a new field $\psi$ by $\beta N(z) = \sqrt{2} \gamma \psi$ with
\begin{align}
    \gamma = \sqrt{\frac{\lambda}{20}\E\left[x^2\right]} \beta,
\end{align}
it can be shown analogously to \cite{Freivogel:2009rf} that all $n$-point functions in this limit reduce to $n$-point functions of free-field vertex operators
\begin{align}
    \Braket{\prod_{j=1}^n {\mathcal O}_{\beta_j}(z_j)} \to \prod_{\substack{i,j=1\\i<j}}^n |z_{i j}|^{4 \gamma_i \gamma_j} = \Braket{\prod_{j=1}^n e^{i \sqrt{2} \gamma_j \psi(z_j)}}.
\end{align}

\section{Conclusions}

Our results  generalize those already obtained  in \cite{Camia:2015ewa, Camia:2019ots} to a very large class of models characterized by the random distribution used to assign values to each loop in computing the value of the layering operator $N$.  The conformal weights of the exponentials of $N$ are determined by the characteristic function of the random distribution, and the correlation functions can be straightforwardly derived using the results of \cite{Camia:2019ots}.

The results raise many further interesting questions.
Our future efforts will focus on understanding the relation between these models and other known 2D conformal field theories. As of now we are aware of only two direct connections. First, the partition function of $n$ free massless bosons can be understood as a BLS with intensity $\lambda = n/2$ \cite{Camia:2015ewa}, and a connection with the free field is discussed in \cite{vandebrug2018} for the winding model on a lattice with $\lambda = 1/2$.
However, the layering operator $N(z)$ considered here (and in \cite{Camia:2015ewa, Camia:2019ots}) is very non-local when expressed in terms of the free field, and it is unclear what the choice of distribution for the random values corresponds to in the free field language.
 
The second connection is to the $n \to 0$ limit of the $O(n)$ model, which can be used to deduce the conformally invariant weights of the ensemble of a single self-avoiding loop in the plane \cite{Gamsa_2006}, and from there, the weights of loops in the BLS and the correlation functions we computed here \cite{Camia:2019ots}. There are no primary operators in the $O(n)$ model with the weights we have computed, so the connection is indirect, but we intend to explore this more deeply in future work.

\appendix
\section{The winding operator} \label{winding}

The winding operator $\nw(z)$ counts the total number of windings of all loops around a point $z$, where each loop is assigned a random orientation. Here, we consider a generalization where each loop is assigned a random value that multiplies the winding number of that loop.

We can extend the analysis of Sec.\ \ref{npointlay_gen} to exponentials of winding-number correlation functions. The main difference from the layering case is that winding number correlation functions factorize into classes of distinct number of windings around each insertion point, since each loop $\gamma$ in an ensemble has an integer number of windings $\theta_\gamma(z_i) = k_i$ around each insertion point $z_i$. 
Extending the derivation as given in \cite{Camia:2015ewa}, we find a formula of the winding correlation functions very similar to \eqref{layeringcorr}:
\begin{align} \label{npointwin}
    \Braket{\prod_{j=1}^n
    e^{i \beta_j N_w(z_j)}
    }_D = &\prod_{\substack{S \subseteq \{z_1, \ldots , z_n \} \\ S \neq \emptyset}} \prod_{K \in \mathbb{Z}^{|S|}} \exp \left[
    - \lambda  \alpha_D(S|S^c;K) \left(1 - \phi\left( \sum_{\substack{i \in I_S}} k_i \beta_i \right) \right) \right].
\end{align}
The weights are given by
\begin{align}
    \alpha_D(S|S^c;K) &= \mul(\gamma: \theta_\gamma(z_i) = k_i ~ \forall z_i \in S, k_i \in K, S^c \in \bar{\gamma}^c, \gamma \in D, \delta \le \operatorname{diam}(\gamma) < R),
\end{align}
with $S \subseteq\{z_1,\ldots,z_n\}$ non-empty and $K$ is a multiset containing one integer $k_i \in \mathbb{Z}$ for each point $z_i \in S$.
For example, the two-point function in some domain $D$ has a contribution from loops that cover $z_1$ with winding number $k_1$ and that do not cover $z_2$ with corresponding weight $\alpha_D(z_1|z_2;k_1)$.

We can use this result to compute the scaling dimensions through the one-point function, analogous to Sec.\ \ref{scalingdims}.
(We remind the reader that it has not been proven that correlation functions of two or more exponentials of winding operators exist in the full plane, but they do exist on any bounded domain \cite{Camia:2015ewa}.)

From \eqref{npointwin} we compute 
\begin{align}
    \Braket{e^{i \beta \nw(z)}}_{\delta,R} = \prod_{k=-\infty}^\infty e^{
    - \alpha_{\mathbb{C}}(z;k) ( 1 - \phi(k \beta) ) },
\end{align}
where loops with zero-winding do not contribute to the one-point function because of \eqref{bounded}.
The loop measure for loops winding around a point for $k \neq 0$ is \cite{Camia:2015ewa, Garban_2006}
\begin{align}
\begin{split}
    \alpha_{\mathbb{C}}(z;k) &= \mul(\gamma: z \in \bar{\gamma}, \theta_\gamma(z)=k, \delta \le \operatorname{diam}(\gamma) < R)\\
	&= \frac{1}{2\pi^2 k^2} \log\frac{R}{\delta},
\end{split}
\end{align}
which leads to
\begin{align}
    \Braket{ e^{i \beta \nw(z)}}_{\delta,R} = \left( \frac{R}{\delta} \right)^{-2 \Delta_\text{w}(\beta)}.
\end{align}
Using property \eqref{herm}, the scaling dimension of winding operator can be written as
\begin{align}
    \Delta_\text{w}(\beta) = \frac{\lambda}{2 \pi^2} \sum_{k=1}^\infty \frac{1}{k^2} (1 - \phi(k \beta)).
\end{align}

\acknowledgments
It is a pleasure to thank Federico Camia and Alberto Gandolfi for discussions. The work of V.\ F.\ is supported by the James Arthur Graduate Award. The work of M.\ K.\ is supported by the NSF through the grant PHY-1820814.

\bibliographystyle{unsrt}
\bibliography{bibliography}

\begin{thebibliography}{10}

\bibitem{Belavin:1984vu}
A.A. Belavin, Alexander~M. Polyakov, and A.B. Zamolodchikov.
\newblock {Infinite Conformal Symmetry in Two-Dimensional Quantum Field
  Theory}.
\newblock {\em Nucl. Phys. B}, 241:333--380, 1984.

\bibitem{Cardy:2008jc}
John Cardy.
\newblock {Conformal Field Theory and Statistical Mechanics}.
\newblock In {\em {Les Houches Summer School: Session 89: Exacts Methods in
  Low-Dimensional Statistical Physics and Quantum Computing}}, 7 2008.

\bibitem{Ginsparg:1988ui}
Paul~H. Ginsparg.
\newblock {Applied Conformal Field Theory}.
\newblock In {\em {Les Houches Summer School in Theoretical Physics: Fields,
  Strings, Critical Phenomena}}, pages 1--168, 9 1988.

\bibitem{DiFrancesco:639405}
Philippe Di~Francesco, Pierre Mathieu, and David Sénéchal.
\newblock {\em {Conformal field theory}}.
\newblock Graduate texts in contemporary physics. Springer, New York, NY, 1997.

\bibitem{Lawler2004}
Gregory~F. Lawler and Wendelin Werner.
\newblock The brownian loop soup.
\newblock {\em Probability Theory and Related Fields}, 128(4):565--588, Apr
  2004.

\bibitem{Camia:2019ots}
Federico Camia, Valentino~F. Foit, Alberto Gandolfi, and Matthew Kleban.
\newblock {Exact Correlation Functions in the Brownian Loop Soup}.
\newblock 2019.

\bibitem{Camia:2015ewa}
Federico Camia, Alberto Gandolfi, and Matthew Kleban.
\newblock {Conformal Correlation Functions in the Brownian Loop Soup}.
\newblock {\em Nucl. Phys.}, B902:483--507, 2016.

\bibitem{camia2015nonbacktracking}
Federico Camia and Marcin Lis.
\newblock Non-backtracking loop soups and statistical mechanics on spin
  networks, 2015.

\bibitem{lejan2010}
Yves Le~Jan.
\newblock Markov loops and renormalization.
\newblock {\em Ann. Probab.}, 38(3):1280--1319, 05 2010.

\bibitem{vandebrug2018}
Tim van~de Brug, Federico Camia, and Marcin Lis.
\newblock Spin systems from loop soups.
\newblock {\em Electron. J. Probab.}, 23:17 pp., 2018.

\bibitem{Freivogel:2009rf}
Ben Freivogel and Matthew Kleban.
\newblock {A Conformal Field Theory for Eternal Inflation}.
\newblock {\em JHEP}, 12:019, 2009.

\bibitem{camia2019brownian}
Federico Camia, Alberto Gandolfi, Giovanni Peccati, and Tulasi~Ram Reddy.
\newblock Brownian loops, layering fields and imaginary gaussian multiplicative
  chaos, 2019.

\bibitem{camia2020limit}
Federico Camia, Yves~Le Jan, and Tulasi~Ram Reddy.
\newblock Limit theorems for loop soup random variables, 2020.

\bibitem{Grimmett_2006}
Geoffrey Grimmett.
\newblock {\em The Random-Cluster Model}.
\newblock Springer Berlin Heidelberg, 2006.

\bibitem{Gamsa_2006}
Adam Gamsa and John Cardy.
\newblock Correlation functions of twist operators applied to single
  self-avoiding loops.
\newblock {\em Journal of Physics A: Mathematical and General},
  39(41):12983–13003, Sep 2006.

\bibitem{werner2005conformally}
Wendelin Werner.
\newblock The conformally invariant measure on self-avoiding loops.
\newblock {\em J. Amer. Math. Soc.}, 21(1):137--169, 2008.

\bibitem{sheffield2009}
Scott Sheffield.
\newblock Exploration trees and conformal loop ensembles.
\newblock {\em Duke Math. J.}, 147(1):79--129, 03 2009.

\bibitem{10.2307/23350642}
Scott Sheffield and Wendelin Werner.
\newblock Conformal loop ensembles: the markovian characterization and the
  loop-soup construction.
\newblock {\em Annals of Mathematics}, 176(3):1827--1917, 2012.

\bibitem{CM_1956-1958__13__76_0}
Eug{\`e}ne Lukacs.
\newblock On certain periodic characteristic functions.
\newblock {\em Compositio Mathematica}, 13:76--80, 1956-1958.

\bibitem{Garban_2006}
Christophe Garban and José A.~Trujillo Ferreras.
\newblock The expected area of the filled planar brownian loop is $\pi/5$.
\newblock {\em Communications in Mathematical Physics}, 264(3):797–810, Mar
  2006.

\end{thebibliography}

\end{document}